# On the Attraction of Matter by the Ponderomotive Miller Force

Rickard Lundin

Swedish Institute of Space Physics, Box 812, 981 28 Kiruna, Sweden

Hans Lidgren

Le Mirabeau, 2 ave des Citronniers, MC 98000, Monaco

rickard.Lundin@irf.se

ABSTRACT

Wave induced attraction of matter is a unique aspect of ponderomotive forcing by electromagnetic (e/m) waves in plasmas. The Miller force, sometimes denoted the gradient force, is of particular interest, because the direction as well as magnitude of the Miller force on a plasma depends on the wave frequency. While plasma is usually considered in its gaseous form, solid bodies can also be treated as plasma, denoted solid-state plasma. The first experimental proof of wave effects in magnetized solid-state plasmas (Lundqvist, 1949, Herlofson, 1950) came after the suggestion by Alfvén (1942) on the possible existence of magneto-hydrodynamic (MHD) waves. However, most of our knowledge basis on MHD/Alfvén waves have since then emerged from space- and laboratory (gaseous) plasmas. It is therefore timely to investigate further the applicability of e/m wave ponderomotive forcing on solid-state plasma.

In this report we discuss the applicability of the ponderomotive Miller force on solid-state plasmas. At this stage the treatise will be rather qualitative, based on theories developed for space plasma. We assume that e/m wave energy and momentum is transferred in a thin surface layer constituting solid-state plasma. A Miller force results from the wave pressure gradient set up in the surface layer. Since the solid body constitute a coupled system (lattice), the Miller force will act on all atoms in the body, the total force limited by the wave input power. A quantitative comparison with experimental results obtained from a Cavendish vacuum experiment (Lidgren and Lundin, 2010) gives a surprisingly good agreement between theory and experiments. As further proof for e/m wave attraction of matter, we show a simulation, taking into account the measured pendulum offsets.

*Subject headings:* Ponderomotive force theory: wave-attraction of matter

1. Introduction

The Miller force (Miller, 1958) is one of at least four fundamental ponderomotive forces governed by electromagnetic waves. The basis of a time-averaged ponderomotive force is that electromagnetic waves transfer energy and momentum to matter. Intuitively, wave-induced ponderomotive forces are expected to be repulsive, i.e. the "heaviness" (in Latin *ponderos*) of waves provides a transfer of energy and momentum in the direction of wave propagation. However, as discussed in a recent review (Lundin and Guglielmi, 2006), ponderomotive forces are more complicated, and have consequences that may contradict intuition. An example of this is magnetic moment pumping, MMP, a ponderomotive force in magnetized plasmas with diverging magnetic fields. The MMP force (Lundin and Hultqvist, 1988, Guglielmi and Lundin, 2001) is always in the direction of the magnetic field divergence - regardless of wave propagation direction and wave frequency. The MMP force exemplifies that the term "heaviness" can be a misleading.

The gradient/Miller force (Guglielmi and Lundin, 2001) is a more striking exception to the intuitive concept of "heaviness" of waves. The direction of the Miller force in plasmas critically depends on the local resonance frequency, e.g. the ion gyrofrequency, $\Omega$. For Alfvén waves in magnetized plasma, the Miller force is repulsive at frequencies above-, and attractive at frequencies below $\Omega$. The Miller force therefore agrees with intuition on wave momentum exchange only in the high-frequency domain. Wave-induced ponderomotive forcing has become an emerging topic in space plasmas (Allan et al., 1990, Li and Temerin, 1993, Shukla et al., 1996, Guglielmi, 1997), for instance with regard to the transfer of energy and momentum by Alfvén waves to plasma. Following the proposal by Alfvén (1942) on the existence of magnetohydrodynamic (MHD) waves, attempts were made to prove the existence of MHD waves in laboratory. A series of experiments carried out by the Royal Institute of Technology in Stockholm proved the existence of the transverse Alfvén mode in magnetized mercury (Lundqvist, 1949, Herlofson, 1950) and magnetized liquid sodium (Lehnert, 1954). While Alfvén waves is a major issue in contemporary space and astrophysical plasma studies, only a few papers on MHD wave propagation in magnetized solid-state plasmas have been published after the proof of existence in the 1950ies, (e.g. Buchsbaum and Galt, 1961, Williams, 1965, Galitis et al., 2000).

However, a number of solid-state effects have been observed connected with microwave radiation. Booske et al. (1998) were the first to associate microwave effects with ponderomotive force mass-transport in solid-state plasmas. In fact, the ponderomotive force referred to by Booske et al (1998) is similar to the force defined by Miller (1958). No doubt ponderomotive e/m wave effects also occur in solid-state.

We will here apply the concept of ponderomotive forcing, specifically the Miller force, to solid-state plasmas. For two reasons, we have chosen Alfvén waves in deriving the Miller force. First, because Alfvén waves have been observed in solid-state plasmas, secondly because they provide a frequency independent response below resonance. Other wave modes may apply, but because solid-state plasma constitutes electrons and holes, instead of electrons and ions, there are limited options.

## 2. The Miller force in plasmas

Consider now Alfvén (magnetohydrodynamic) waves in plasma of ions and electrons. Since the ion mass is typically >1800 times higher than the electron mass, the electron mass can be neglected. The mass density and the corresponding forcing upon the plasma is therefore determined by the ion mass, $m$. Alfvén waves propagate along magnetic field lines ($\mathbf{k}=(0,0,k)$) and have linear polarization

$\mathbf{E} = (1,0,0) E \exp(-i\omega t)$, $\mathbf{b} = (0,1,0) b \exp(-i\omega t)$. Hence $\mathbf{v} = (i\omega/\Omega, 1, 0) v \exp(-i\omega t)$, where

$$v = \frac{e\Omega E}{m(\omega^2 - \Omega^2)}, \qquad (1)$$

$\omega$ is the wave frequency, $e$ is the electronic charge, and $\Omega$ is the ion cyclotron frequency of the plasma. This leads to the following expression of the longitudinal *gradient or ponderomotive Miller force* driven by Alfvén waves (Alfvén, 1942):

$$F_z = -\frac{e^2}{4m(\omega^2 - \Omega^2)} \frac{\partial E^2}{\partial z}. \qquad (2)$$

Notice that the expression has a singularity at $\omega = \Omega$. Moreover, the Miller force is attractive for $\omega < \Omega$ and repulsive for $\omega > \Omega$. The obvious implication of this is that low-frequency Alfvén waves ($\omega < \Omega$) will attract ions towards the wave source, while high-frequency Alfvén waves will repel ions. We also note that the Miller force does not depend on the sign of the charge. The fact that electromagnetic waves in plasma may deliver momentum "backward" (-z), i.e. leading to attraction, is not obvious. However, MHD waves are a class of media waves. Plasma and magnetic field display a mutual oscillation, the plasma considered "frozen" into the magnetic field. The local resonance frequency, rather than the direction of wave propagation (+z-direction) determines the direction of the force. Qualitatively a wave frequency much lower than the gyrofrequency implies that ions and electrons experience a quasi-static wave electric field, the differential drift of electrons and ions leading to a depolarization electric field. The opposing depolarization field (E*) gives a force in the opposite direction to the incident wave. Notice that the electric field gradient $(\partial E^2/\partial z)$ determines the magnitude of the force.

From $\Omega = eB/mc$ we have in the low frequency limit $\omega^2 \ll \Omega^2$:

$$F_z = \frac{mc^2}{2B^2}\frac{\partial E^2}{\partial z} \qquad (3)$$

This implies that the force is constant and independent of wave frequency in a homogeneous medium ($B$ = constant) in the low frequency limit, the force being proportional to the gradient of the EM-wave intensity $(\partial E^2/\partial z)$. Because the wave intensity is decreasing while interacting with matter (exerting force on matter) the force $F_z$ is directed opposite to the wave propagation direction.

### 3. Miller force on solid bodies

The concept of MHD waves stems from the fluid description of plasmas. Hannes Alfvén published the theory of MHD waves in 1942, but it took seven years to prove their existence, and then in magnetized solid-state plasmas (Lundqvist, 1949, Herlofson, 1950, Lehnert, 1954). MHD waves are in essence medium waves governed by the magnetic tension in magnetized plasma - the stronger the magnetic tension, the weaker the wave group velocity and Miller force. In a similar way, medium waves in solids are governed by interatomic tension. While the local resonance frequency in gaseous plasma is determined by the ion gyrofrequency, the local resonance frequency in solid-state plasma (electrons and holes) is less obvious. A simple approach is to consider the interatomic distance in the framework of lattice vibrations, here using the wave description instead of the particle (phonon) description.

Under the assumption that e/m waves irradiating solid-state plasma are linearly polarized, producing a media-specific response, we may use the Miller force theory (expression 2) from MHD. In so doing, we replace the cyclotron frequency with the media specific (interatomic) resonance frequency, $\Omega_a$. Incident e/m waves may propagate until stopped/absorbed, but due to the strong atomic binding force the Miller force will be distributed and affect all atoms in the lattice. The force (in cgs units) exerted on individual particles/atoms of mass $m_a$ by linearly polarized waves with a radiation electric field $E$, is:

$$F_z = -\frac{e^2}{4m_a(\omega^2 - \Omega_a^2)}\frac{\partial E^2}{\partial z} \quad (4)$$

Fig. 1 illustrates the theoretical Miller force normalized to the resonance frequency, $\Omega_a$. The Miller force is independent of wave frequency for $\omega^2 \ll \Omega_a^2$ and attractive over the entire frequency range below resonance. The Miller force is repulsive at frequencies above resonance, but decays strongly at higher frequencies.

The resonance cavity and the corresponding resonance frequency, $\Omega_a$ is determined by the inter-atomic separation distance, $a$. For frequencies well below resonance, $\omega^2 \ll \Omega_a^2$, we arrive at the following expression for the Miller force:

$$F_z = \frac{a^2 e^2}{4mc^2}\frac{\partial E^2}{\partial z} = \xi(a,m)\frac{\partial E^2}{\partial z} \quad (5)$$

The force depends on a material constant, $\xi(a,m)$, and the spatial gradient (damping) of the wave electric field propagating into matter. No damping implies no forcing, i.e. either total reflection or unobstructed propagation through the media. Immediate and full damping implies maximum forcing at a thin surface layer of the body.

To analyze the energy and momentum transfer we start by assuming that all the wave energy goes into Miller forcing, bearing in mind that some fraction of the wave energy goes into heating and the repulsive radiation force. We then define a penetration depth, i.e. $E$ may penetrate a certain depth $d(\lambda)$, where $\lambda$ is the wavelength of the incident radiation. Assuming for the sake of simplicity that $d(\lambda) \approx a \cdot \exp(1/(1-a/\lambda))$. A material constant $\xi(a,m)$ and a gradient $\delta E^2 / \delta z$ related with $d(\lambda)$ and the transverse electric field $E$ of incident waves determines the force exerted on matter.

We may now test the analytical expression (4) against results from the Cavendish experiment performed in vacuum (Lidgren and Lundin, 2010). The lead target sphere has a mass of 20 g and is illuminated by a lamp (26 W) covered by aluminium at a distance of 1.5 cm (Lidgren and Lundin, 2010). The effective lead-target area is 0.87 cm$^2$. Isotropic radiation in infrared (IR) from the lamp is here anticipaped shortly after power-on. The distance from the light source and the lead sphere target area determines the net irradiance on the lead sphere. After power-off, the real lamp continues to radiate (in IR) with exponential decaying intensity. However, we will here first assume that the lamp goes directly into full power, and shuts off instantly with no decay.

Because a solid body constitutes a coupled system of atoms, wave forcing will affect the entire body regardless of the wave penetration depth. The force may act on a thin target layer, but the net force will act on the total mass. Miller forcing is now estimated on basis of the following data: The number of lead atoms in the target sphere is estimated to $5.7 \cdot 10^{22}$ atoms (target mass 20 g, $\rho = 11.4$ g/cm$^{-3}$). From this and the target volume, the inter-atomic distance, $a$, is estimated to $\approx 2 \cdot 10^{-8}$ cm. Inserting the above values in expression (5), the net force on all lead atoms becomes:

$$F \approx 4.18 \cdot 10^{-12} \frac{\partial E^2}{\partial z} \quad (dyn) \qquad (6)$$

What remains to be determined is the gradient of the EM-wave intensity $(\partial E^2/\partial z)$ in the conducting (lead) material within which the primary forcing takes place. Considering the expression for $d(\lambda)$ and the wavelength $\lambda$ for visible light, we get $\Delta z \approx 2.7 \cdot a$. With all wave energy absorbed by the target and subsequently converted to Miller forcing, we get an equivalent force on the body:

$$F \approx 4.18 \cdot 10^{-12} \frac{\Delta E^2}{\Delta z} \approx 7.83 \cdot 10^{-5} E^2 \quad (dyn) \tag{7}$$

Assuming isotropic radiation, the fraction reaching the spherical target is estimated to $I \approx 0.72$ W. This gives a radiation pressure force ($F_{rad} = I_{targ}/c$) of $F_{rad} \approx 2.4 \cdot 10^{-9}$ N, to be compared with the Miller force. From the relation between the rate of incident radiation energy ($I$) and the electric field ($E$) of the EM-wave ($I = 0.5 \cdot c \cdot \varepsilon_0 \cdot E^2$; in SI units), we get $I \approx 1.33 \cdot 10^{-3} \cdot E^2$ (W/m$^2$). Inserting into expression (7) we get in cgs units: $F = 3.8 \cdot 10^{-4}$ (dyn), or correspondingly in SI-units $F = 3.8 \cdot 10^{-9}$ (N).

In the above calculation we have assumed that the entire irradiated 0.72 W is transferred to Miller forcing. This represents an extreme with no heat production, no reflectance and no emissivity of the target. Furthermore, the Miller force derived is 57% higher than the theoretical radiation pressure force ($F_{rad}$). From a theoretical point of view, identical forces may be obtained by adjusting the wave penetration depth to $\Delta z \approx 4.2 \cdot a$. However, this analysis contains a number uncertainties about geometries, lamp power etc. A more detailed description on penetration depths lies outside the scope of this paper. What is of interest here is that the value of the Miller force is close to the experimental value, $2.9 \pm 0.6 \cdot 10^{-9}$ (N), determined by Lidgren and Lundin (2010).

An important aspect of the Miller forcing is how the net attraction works. Although acting on individual atoms, it becomes most important in a collective sense when sufficient mass is involved. This is why the radiation pressure force first presented by Maxwell in 1871, and first tested by Lebedeev (1901), leads to a repulsive force only for thin and low mass target materials. A *gedanken* experiment illustrates this. Replacing the solid lead ball with a 0.5 mm thin lead foil with similar diameter as the lead ball, the mass of the lead foil will be 1.0 g, and the number of atoms $2.9 \cdot 10^{21}$. Irradiation by the same wave power leads to an attractive Miller force corresponding to $1.9 \cdot 10^{-10}$ (N), about 1/10 of the force exerted by the nominal radiation pressure ($F_{rad} \approx 2.4 \cdot 10^{-9}$ N). An experiment carried out with a thin foil is therefore expected to lead to a repulsive force. In our opinion, this was the main reason why thin foils were required to prove of existence of the light pressure force (Lebedeev, 1901). Using thin foils Lebedeev arrived to some 80% of the predicted light pressure. In our gedanken experiment above, assuming that 10% of the energy goes into isotropic heating, we reach a value that is 82% of the theoretical radiation pressure repulsive force, in fair agreement with the Lebedeev results.

As a final test of the validity of the theory we have made an analytic simulation of the Cavendish pendulum experiment based on the input values used in the V2-experiment (Lidgren and Lundin, 2010). Notice that the radiation from the lamp in the V2-experiment illuminated the lead ball over an extended period of time (up to one hour). The radiation from the lamp emerged as infrared, decaying according to the empiric expression (Fig. 4 in Lidgren and Lundin. 2010):

$$I(t) = I_0 \exp(-at) \quad t \geq t_0 \quad (W) \quad (8)$$

Where $I_0$ is the peak lamp power and *(a)* is a constant, where *a=0,0011* for experiment V2. From the above expression of the lamp power we can compute the force versus time using expression (7), and subsequently simulate the displacement caused by the wave forcing. In so doing we also take into account the geometrical effect caused by the pendulum displacement, i.e. the change of forcing with distance between the lamp and the lead ball. For the sake of simplicity we assume that the radiation decrease with distance, r, as $1/r^2$. The result is shown in Fig. 2, the upper panel displaying the experimental results, the lower panel the simulation. Power on of lamp is marked with black dot. The pendulum displacement, ΔS on the Y-axis, gives the actual displacement (true motion) of the pendulum. To enable a detailed comparison of the displacement between experiment and simulations, we have multiplied the displacement amplitude by 1.15 in the simulation panel. The simulated response on basis of the total force distributed according to expression (8) gave an underestimate of the displacement values by 15% compared to the measured values. Nevertheless, the overall agreement between experiment and model is good. Notice in particular the exponential decrease of the displacement offset, in the simulation determined from the exponentially decaying radiation force affecting the torsion pendulum torque.

## 4. Summary

We have demonstrated that ponderomotive forcing by electromagnetic waves is capable of causing attraction of solid bodies. The ponderomotive Miller force may be feeble for individual atoms, compared to the Maxwell force (radiation pressure), but it becomes of major importance for massive bodies where the net attraction force is collective, acting on all atoms in a body. The Miller force is induced by waves penetrating a body, the force being proportional to the spatial gradient of the square of the wave $E$-field (expressions 5-7). The latter indicates that Miller forcing is most effective for conducting bodies. Experimental measurements in high vacuum (Lidgren and Lundin (2010) offer unambiguous proof for electromagnetic wave-induced attraction of matter. The Miller force theory leads to estimates that are in general agreement with experimental results ($\approx 2.9 \pm 0.6 \cdot 10^{-9}$ N in the vacuum experiment). A simulation of the experimental findings could also be made on basis of the Miller force theory and a decaying radiation source/lamp. With a decaying radiation from a lamp, and a corresponding decaying Miller force, we can replicate the overall characteristics of the Cavendish pendulum motion in vacuum.

The physical implications of the theory proposed, and the experimental results (Lidgren and Lundin, 2010) are fairly obvious. On a macro-scale, it is expected to have implications for space probes and satellites orbiting the Earth, in the latter case when going in and out of eclipse. On a micro-scale the Miller force resembles at first glance the Casimir-force, first predicted by Casimir (1948), later confirmed by Lamoreaux (1997). However, the Casimir force stems from quantum electrodynamic considerations of vacuum fluctuations, whereas the Miller force on matter is a direct consequence of electromagnetic radiation.

REFERENCES

Alfvén, H., Nature, 1942, 150, 405

Allan, W., J. R. Manuel, and E. M. Poulter, Geophys. Res. Lett., 1990, **17**, 917

Booske, J.H., R-F. Cooper, S.A. Freeman, K.I. Rybakov, and V.E. Semenov, Physics of Plasmas, 1998, 5, 5, 1664-1670

Buchsbaum, S. J.; Galt, J. K., Physics of Fluids, 1961, Vol. 4, p.1514-1516

Casimir, H. B. G., Koninkl. Ned. Akad. Weteschap. Proc., 1948, 51, 793

Galitis A., Lielausis O., Dementev S., et al., 2000, Phys Rev Lett, 84, 4365

Guglielmi, A. and R. Lundin, J. Geophys. Res, 2001, 106, 13219

Guglielmi A., 1997, J. Geophys. Res., **102**, 209

Herlofson, N., 1950, Nature, 165, 1020.

Lamoreaux, S. K., Phys. Rev. Lett., 1997, **78**, 5

Lehnert, B., Phys. Rev, 1954, 76, 1805

Lebedev, P., Annalen der Physik, 1901

Li, X., and M. Temerin, Geophys. Res. Lett., 1993, **20**, 13

Lidgren, H., and R. Lundin, arXiv subm., 2010.

Lundin, R. and A. Guglielmi, Space Sci. Rev, 2006, DOI 10.1007/s11214-006-8314-8

Lundin, R. and B. Hultqvist, J. Geophys. Res., 1988, 94, 6665

Lundqvist, S., 1949, Phys. Rev., 76, 1805

Miller, M. A., Radiophysics (Russia), 1958, **1**, 110

Shukla P. K., L. Stenflo, R. Bingham, and R. O. Dendy, 1996, J. Geophys. Res., **101**, 27449

Williams, G. A., 1965, Phys. Rev., 139, 3a, 771-778, 1965

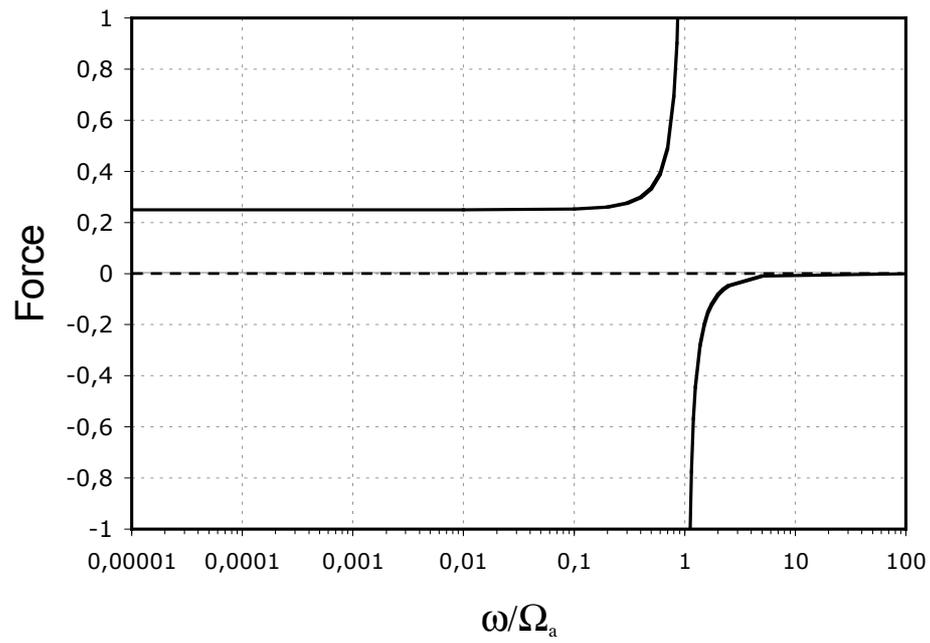

Figure 1 — Miller force versus normalized frequency for unit material constant ($e^2/4m_a$), and $\partial E^2/\partial z = 0.25$ (expression (4)).

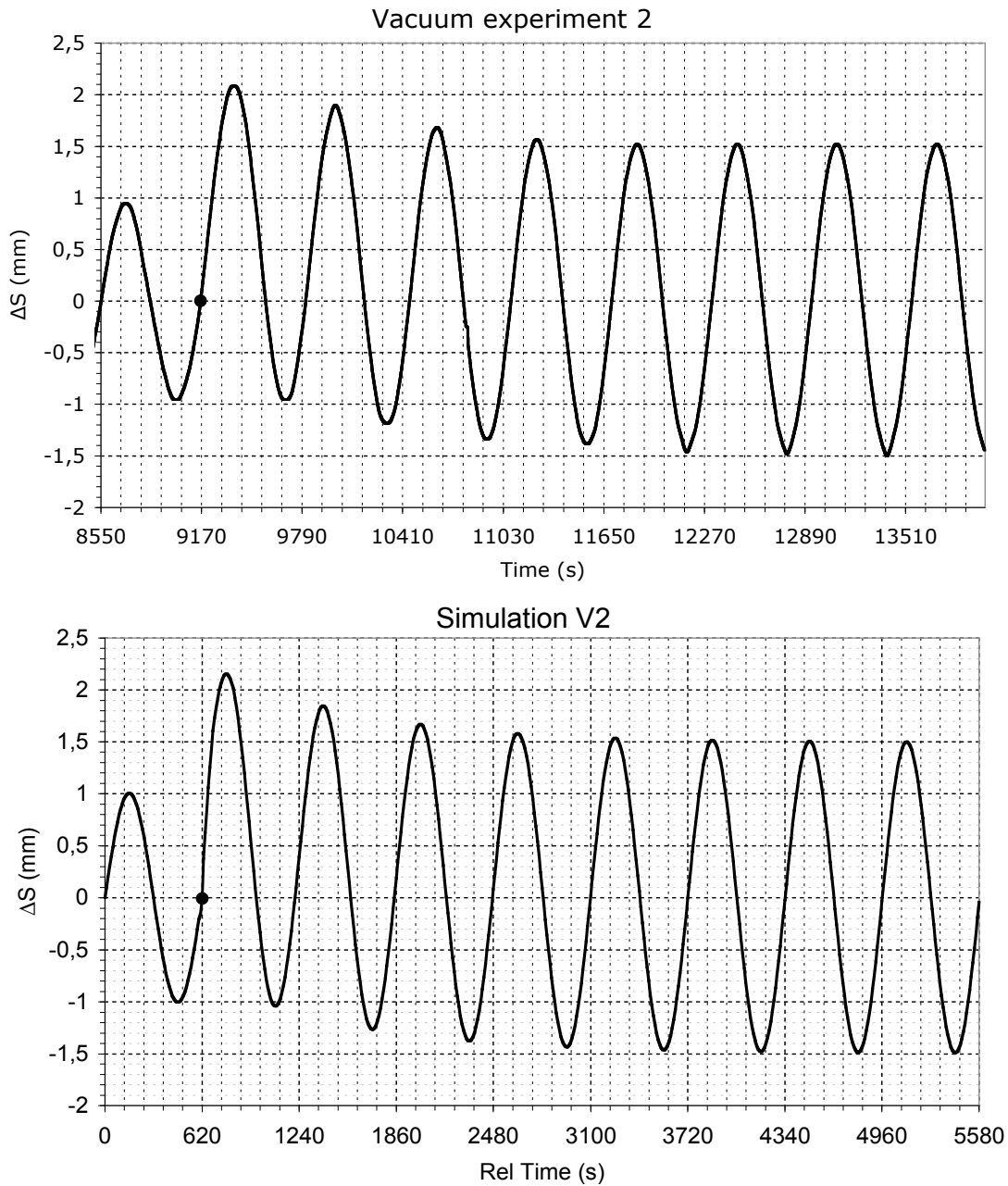

Figure 2. —Pendulum motion of Cavendish experiment in vacuum. An exponentially decreasing attractive force is applied to one of the lead balls in the torsion pendulum experiment. Black dot marks lamp power-on.

Upper panel: Pendulum motion in vacuum experiment V2 (Lidgren and Lundin, 2010).

Lower panel: Simulated pendulum displacement of the Cavendish experiment in vacuum.